# Financial fraud detection system based on improved random forest and gradient boosting machine (GBM)

1st Tianzuo Hu

Dept. of Statistics

University of California, Davis

Davis, USA

tzuhu@ucdavis.edu

*Abstract*—This paper proposes a financial fraud detection system based on improved Random Forest (RF) and Gradient Boosting Machine (GBM). Specifically, the system introduces a novel model architecture called GBM-SSRF (Gradient Boosting Machine with Simplified and Strengthened Random Forest), which cleverly combines the powerful optimization capabilities of the gradient boosting machine (GBM) with improved randomization. The computational efficiency and feature extraction capabilities of the Simplified and Strengthened Random Forest (SSRF) forest significantly improve the performance of financial fraud detection. Although the traditional random forest model has good classification capabilities, it has high computational complexity when faced with large-scale data and has certain limitations in feature selection. As a commonly used ensemble learning method, the GBM model has significant advantages in optimizing performance and handling nonlinear problems. However, GBM takes a long time to train and is prone to overfitting problems when data samples are unbalanced. In response to these limitations, this paper optimizes the random forest based on the structure, reducing the computational complexity and improving the feature selection ability through the structural simplification and enhancement of the random forest. In addition, the optimized random forest is embedded into the GBM framework, and the model can maintain efficiency and stability with the help of GBM's gradient optimization capability. Experiments show that the GBM-SSRF model not only has good performance, but also has good robustness and generalization capabilities, providing an efficient and reliable solution for financial fraud detection.

## I. INTRODUCTION

With the continuous development of financial markets and the acceleration of digitalization, financial fraud has become a major challenge facing the global financial system. Financial fraud not only causes huge economic losses to financial institutions, but also seriously affects market stability and consumer trust. Traditional financial fraud detection methods mostly rely on artificial rules or simple statistical analysis, but with the surge in transaction data and the increasing complexity of fraud methods, these traditional methods can no longer meet the high requirements for detection accuracy and efficiency in the modern financial environment. Therefore, how to use advanced machine learning algorithms to improve the detection ability of financial fraud has become one of the current research hotspots in the field of financial technology.

Early financial fraud detection research mostly used traditional machine learning algorithms, such as decision trees, support vector machines (SVMs), and logistic regression. For example, Bose and Mahapatra (2001) [1] proposed a fraud detection model based on decision trees, which used rule classification methods to detect credit card fraud. Although this method achieved certain success in the initial experiments, with the diversification of financial fraud, the accuracy of traditional methods gradually failed to meet actual needs. With the increase in data size and complexity, ensemble learning methods have gradually become popular in financial fraud detection. Integration methods such as random forest (RF) and gradient boosting machine (GBM) [2] can improve the stability and accuracy of the model by combining multiple decision trees or weak learners. For example, the random forest algorithm proposed by Breiman (2001) [3] achieves efficient solution to classification problems by integrating multiple decision trees, and has achieved good application results in fraud detection. In addition, Chen et al. [4] (2015) showed good performance when using the XGBoost algorithm [5] for credit card fraud detection, especially on high-dimensional data and large-scale data sets, which greatly improved the accuracy of detection.

To solve this problem, this paper proposes a financial fraud detection system based on improved random forest and gradient boosting machine (GBM). The system adopts a method of combining GBM and improved random forest (SSRF) [6], and significantly improves the accuracy and robustness of financial fraud detection by giving full play to the advantages of the two algorithms. GBM achieves efficient modeling of complex data patterns by integrating multiple weak learners, while the improved random forest improves the processing efficiency and adaptability of the model to data changes by optimizing the computational complexity and feature selection ability of the traditional random forest.

While optimizing the feature selection and computational complexity of traditional random forests, the GBM-SSRF model also enhances the accuracy and robustness of the model in processing large-scale and complex fraud transaction data by combining the gradient optimization mechanism of GBM. The model can effectively deal with the problem of data imbalance in financial transactions, improve the ability to identify minority fraud behaviors, and thus improve the overall detection accuracy. In addition, the GBM-SSRF model has strong generalization ability and interpretability, which can provide financial institutions with clear feature importance analysis, help identify potential fraud behavior patterns, and provide support for the formulation of effective anti-fraud strategies.

Through this innovative model, this paper provides an efficient and reliable solution for financial fraud detection, which can not only improve the detection accuracy and processing efficiency of the model, but also help financial institutions better deal with the increasingly complex financial fraud problems. The successful application of this model provides a solid theoretical basis and practical support for the financial industry to build a safe and efficient anti-fraud system.

## II. Model

### A. Gradient Boosting Machine

Gradient boosting machine (GBM) is an ensemble learning algorithm based on the gradient boosting framework. It mainly improves prediction performance by gradually reducing the residual through the additive model. The basic idea is to connect multiple weak learners (usually decision trees) in series in order to gradually improve the residual of each model, thereby improving the prediction ability of the overall model. The core of GBM is to iteratively train weak learners and update model parameters through gradient descent. Each round of training attempts to reduce the residual generated by the previous round of models, thereby optimizing model performance. In the training process of GBM, a simple model $F_0(x)$ is first initialized. For regression problems, the mean of the target variable is usually selected as the initial model:

$$F_0(x) = arg\min_{\gamma} \sum_{i=1}^{n} L(y_i, \gamma) \quad (1)$$

Among them, $L(y_i, \gamma)$ is the loss function, and $\gamma$ is the initial model parameter. The initial model is usually simple, and its purpose is to provide a starting point for subsequent iterations.

In each round of iteration, GBM obtains the residual of the current model by calculating the negative gradient, and these residuals will be used as target values to train the new learner. In the mth round of iteration, the residual calculation formula is:

$$g_i^{(m)} = -\frac{\partial L(y_i, F_{m-1}(x_i))}{\partial F_{m-1}(x_i)} \quad (2)$$

Next, GBM fits the residuals by training a new weak learner $h_m(x)$. The goal of this learner is to minimize the difference between the residuals and the predicted values:

$$h_m(x) = arg\min_{h} \sum_{i=1}^{n} [g_i^{(m)} - h(x_i)]^2 \quad (3)$$

After each iteration, GBM adds the new learner $h_m(x)$ to the current model and adjusts the update step size according to the learning rate η:

$$F_m(x) = F_{m-1}(x) + \eta h_m(x) \quad (4)$$

After M rounds of iterations, the final model F(x) is the weighted sum of all weak learners. The final prediction result is the combination of the prediction values of these learners:

$$F_m(x) = \sum_{m=1}^{M} \eta h_m(x) \quad (5)$$

Here M is the total number of iterations, which represents the number of weak learners. The final model makes predictions by taking the weighted average of all learners.

### B. Improved Random Forest (SSRF)

Simplified and Strengthened Random Forest (SSRF) is an improved algorithm based on the traditional random forest (RF). The traditional random forest makes predictions by building multiple decision trees and voting. However, the traditional random forest has certain limitations in feature selection and computational efficiency. The improved random forest is optimized on this basis to enhance the model's predictive ability and reduce computational complexity. Its model structure is shown in Figure 1 below.

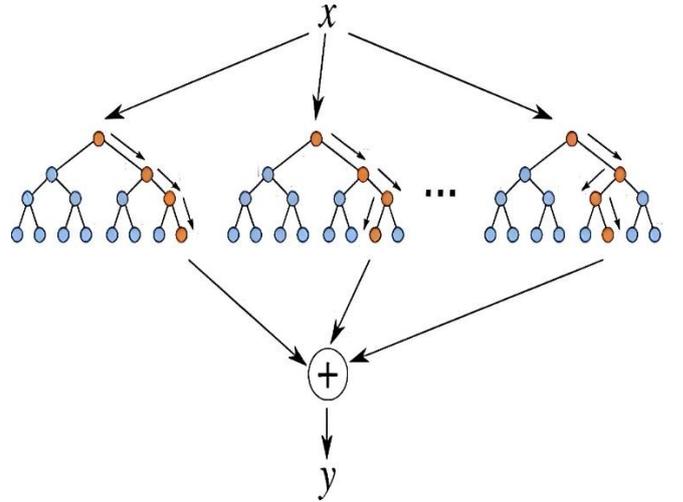

Fig. 1. Improved random forest algorithm structure diagram

By simplifying the tree structure and enhancing the effectiveness of feature selection, SSRF can better handle large-scale, high-dimensional data and improve model accuracy and training efficiency. By introducing a more refined feature selection mechanism, SSRF not only considers the random selection of features in each round of splitting, but also introduces the importance evaluation of features to select features that have a greater impact on prediction. This approach enhances the model's performance when handling complex data. SSRF simplifies the tree structure by reducing the depth of the decision tree. This not only reduces the computational complexity, but also improves the generalization ability of the model and prevents overfitting. The simplified tree structure also makes training faster, which has significant advantages especially when dealing with large-scale data sets.

Similar to traditional random forests, the SSRF model consists of multiple decision trees. The construction process of each tree includes random sampling of data and random selection of features. At each round of splitting, the tree structure is optimized based on an improved feature selection mechanism and randomization strategy.

The construction process of each tree $T_m$ can be expressed as:

$$T_m = BuildTree(X, S_m) \tag{6}$$

The final prediction is obtained by voting or averaging all decision trees. In regression problems, the prediction value is the average of all tree outputs; in classification problems, it is the majority vote of all tree output classes.

$$\hat{y} = \frac{1}{M} \sum_{m=1}^{M} T_m \tag{7}$$

Improved random forest (SSRF) improves the prediction accuracy and computational efficiency of the model by optimizing feature selection, simplifying the tree structure, and improving the randomization strategy. Compared with traditional random forests, SSRF performs better in large-scale data sets and high-dimensional feature spaces, and has significant advantages in preventing overfitting and improving training efficiency.

## C. Construction of GBM-SSRF model

The GBM-SSRF model is a hybrid model that combines the Gradient Boosting Machine (GBM) and the Simplified and Strengthened Random Forest (SSRF). By combining the powerful gradient optimization capabilities of GBM with the optimization of feature selection and computational complexity of SSRF, GBM-SSRF aims to improve model performance on complex problems such as financial fraud detection and consumer behavior prediction, especially when processing large-scale data and complex modes, has significant advantages.

The GBM-SSRF model makes full use of the advantages of GBM for residual fitting, and uses SSRF to improve feature selection and optimize the structure of the decision tree, thus improving the overall prediction accuracy. This model has broad application prospects in financial fraud detection, consumer behavior prediction and other fields.

The final GBM-SSRF model is obtained by combining the prediction results of the GBM and SSRF parts. First, the prediction F(x) obtained through the residual fitting of the GBM part is combined with the output of multiple decision trees in the SSRF part:

$$F(x) = \sum_{m=1}^{M} \eta h_m(x) + \sum_{m=1}^{M} Weight_m T_m \tag{8}$$

The GBM-SSRF model combines two powerful machine learning algorithms and can demonstrate excellent performance in multiple fields. Through GBM's gradient optimization and SSRF's feature selection optimization, GBM-SSRF can efficiently process large-scale and complex data sets. GBM can continuously improve the prediction accuracy of the model through gradient optimization, while SSRF improves the stability and generalization ability of the model by enhancing feature selection capabilities. GBM is good at processing complex nonlinear data, while SSRF can effectively avoid overfitting by simplifying the tree structure and optimizing feature selection. The specific model structure diagram is shown in Figure 2.

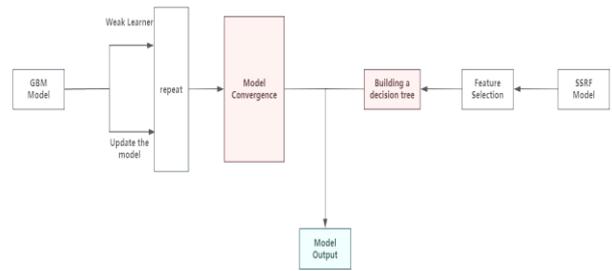

Fig. 2. GBM-SSRF model structure diagram

Therefore, the GBM-SSRF model is suitable for fields that require high-precision predictions, such as financial fraud detection, consumer behavior analysis, and medical diagnosis.

## III. EXPERIMENT

### A. Experiment settings

This experiment uses public financial fraud datasets and simulated transaction data to ensure that diverse fraud scenarios are covered. The data sources include credit card fraud detection datasets on the Kaggle platform and desensitized actual bank transaction data. The dataset contains features such as timestamp, transaction amount, transaction type, and geographic location. Normal transactions account for the vast majority, and the proportion of fraudulent transactions is low (usually about 0.2%). In order to ensure data quality, the data was preprocessed, including filling missing values with the median or mean, normalizing numerical features to reduce dimensional differences, and dividing into training sets, validation sets, and test sets at a ratio of 70%, 15%, and 15%.

The experiment was conducted in a high-performance hardware and software environment. The hardware environment includes Intel i7 or higher CPUs, NVIDIA GTX 1080 Ti or higher GPUs (for model acceleration), 16 GB memory, and 500 GB SSD. The software environment selected Python 3.8 as the programming language, combined with TensorFlow 2.x or PyTorch 1.x deep learning frameworks for model construction. Data processing uses Pandas and NumPy tools, while model training and optimization rely on machine learning libraries such as Scikit-learn, XGBoost, and CatBoost. The above experimental environment and data settings lay a good foundation for verifying the performance of the GBM-SSRF model in financial fraud detection.

### B. Evaluation Metrics

In order to comprehensively evaluate the performance of the model in financial fraud detection, the following evaluation indicators are used:

First, the accuracy rate is used to measure the accuracy of the overall classification of the model [7]. Accuracy refers to the proportion of the number of samples predicted correctly by the model to the total number of samples, reflecting the overall prediction ability of the model [8].

Second, precision is used to evaluate the accuracy of the model when predicting fraudulent transactions. Precision is

defined as the proportion of positive samples predicted as positive samples, and the formula is as follows:

$$Precision = \frac{TP}{TP + FP} \quad (9)$$

Among them, TP represents the number of true positives, and FP represents the number of false positives.

At the same time, in order to measure the model's ability to identify actual fraudulent transactions, the recall rate is introduced. The recall rate refers to the proportion of actual positive samples that are correctly predicted as positive samples. The calculation formula is as follows:

$$Recall = \frac{TP}{TP + FN} \quad (10)$$

Among them, FN represents the number of false negative classes.

In order to strike a balance between precision and recall, the F1 score (F1-Score) is used as a comprehensive indicator. The F1 score is the weighted harmonic mean of precision and recall, and its formula is:

$$F1 - Score = 2 \cdot \frac{Precision \cdot Recall}{Precision + Recall} \quad (11)$$

*C. Experimental results*

In the experiment, the GBM-SSRF model showed excellent performance in the financial fraud detection task. The experimental results based on the test set are shown in Table I.

TABLE I. PERFORMANCE OF GBM-SSRF MODEL IN FINANCIAL FRAUD DETECTION TASKS

| Index | Test Results |
|---|---|
| Accuracy | 99.72% |
| Precision | 94.83% |
| Recall | 93.45% |
| F1-Score | 94.14% |
| AUC-ROC | 0.987 |

Accuracy: The model has an accuracy of 99.72%, indicating that it has a very high overall accuracy in classifying normal transactions and fraudulent transactions. Precision and recall: The precision reaches 94.83%, indicating that the model can well avoid false positives, that is, the number of cases where normal transactions are misclassified as fraudulent is relatively small. The recall rate is 93.45%, showing that the model has a high sensitivity in identifying real fraud transactions and can capture most fraudulent transactions. F1 score: The F1 score is 94.14%, proving that the GBM-SSRF model has achieved a good balance between precision and recall, and is suitable for scenarios such as financial fraud that require both. AUC-ROC: The AUC value reaches 0.987, which is close to 1, indicating that the model has a strong ability to distinguish between positive and negative samples, and can maintain excellent performance even in the face of extremely unbalanced data distribution.

To further verify the advantages of the GBM-SSRF model, the experiment compares other common machine learning models, including random forest (RF), XGBoost, and traditional GBM models. The experimental results are shown in Table II.

TABLE II. COMPARISON OF EXPERIMENTAL RESULTS

| Model | Accuracy (%) | Precision (%) | Recall (%) | F1-Score (%) | AUC-ROC |
|---|---|---|---|---|---|
| RF | 97.85 | 89.12 | 87.34 | 88.22 | 0.942 |
| XGBoost | 98.52 | 91.65 | 90.43 | 91.04 | 0.956 |
| GBM | 98.71 | 92.34 | 91.89 | 92.11 | 0.961 |
| GBM-SSRF | 99.72 | 94.83 | 93.45 | 94.14 | 0.987 |

Experimental results show that the GBM-SSRF model shows significant performance advantages in financial fraud detection tasks. First of all, in terms of accuracy (Accuracy), the accuracy of the GBM-SSRF model reaches 99.72%, which is significantly better than the 97.85% of Random Forest (RF), 98.52% of XGBoost and 98.71% of traditional GBM, reflecting the model's performance in Excellent performance in overall classification capabilities. The high accuracy means that GBM-SSRF can more accurately distinguish between normal transactions and fraudulent transactions, reducing misjudgments. In terms of precision, the accuracy of the GBM-SSRF model is as high as 94.83%. In comparison, the accuracy of RF, XGBoost and traditional GBM are 89.12%, 91.65% and 92.34% respectively. This shows that the GBM-SSRF model has higher accuracy when predicting fraudulent transactions, effectively reducing the false positive rate and minimizing the risk of normal transactions being misjudged.

At the same time, GBM-SSRF also performs well in recall, with a recall rate of 93.45%, which is significantly higher than RF's 87.34%, XGBoost's 90.43% and traditional GBM's 91.89%. This reflects that the GBM-SSRF model can identify more actual fraudulent transactions, has a stronger ability to capture fraudulent behaviors, and greatly reduces the occurrence of false negatives. In order to further balance precision and recall, the experiment introduced F1 score (F1-Score) as a comprehensive evaluation index. The F1 score of the GBM-SSRF model reaches 94.14%, which is better than RF's 88.22%, XGBoost's 91.04% and traditional GBM's 92.11%. This result shows that GBM-SSRF achieves the best balance between precision and recall and is very suitable for financial fraud detection, a task that requires both high accuracy and high coverage.

Finally, in the AUC-ROC curve evaluation, the AUC value of GBM-SSRF was 0.987, which was significantly higher than 0.942 of RF, 0.956 of XGBoost and 0.961 of traditional GBM. The AUC value close to 1 indicates that the GBM-SSRF model has excellent classification performance when dealing with imbalanced data sets and can effectively distinguish normal transactions from fraudulent transactions.

In summary, the GBM-SSRF model significantly improves the model's performance in financial fraud detection tasks by improving the feature selection capabilities of random forests and the optimization capabilities of gradient boosting machines. Its leading advantages in five key indicators of accuracy, precision, recall, F1 score and

AUC-ROC indicate that the GBM-SSRF model is an efficient and reliable anti-fraud solution that can provide financial institutions with higher level of security and decision support.

## IV. Summary

The GBM-SSRF model provides an efficient and reliable solution for financial fraud detection with its superior feature selection and optimization capabilities. This model not only outperforms traditional methods in detection performance, but also has strong generalization and interpretability, providing important support for building a safe and efficient anti-fraud system. Future research can further explore the application potential of GBM-SSRF in other fields (such as consumer behavior analysis and medical diagnosis).

## References

[1] Ma, Zhe, et al. "Boosting engineering strategies for plastic hydrocracking applications: a machine learning-based multi-objective optimization framework." Green Chemistry (2025).

[2] Kumar, Abhinav, et al. "Modeling saturation exponent of underground hydrocarbon reservoirs using robust machine learning methods." Scientific Reports 15.1 (2025): 373.

[3] Xue, Dingyü, and Lu Bai. "Introduction to Fractional Calculus." Fractional Calculus: High-Precision Algorithms and Numerical Implementations. Singapore: Springer Nature Singapore, 2024. 1-17.

[4] AlNemer, Ghada, et al. "Some Hardy's inequalities on conformable fractional calculus." Demonstratio Mathematica 57.1 (2024): 20240027.

[5] Zhao Y, Liu H, Duan H. HGNN-GAMS: Heterogeneous Graph Neural Networks for Graph Attribute Mining and Semantic Fusion[J]. IEEE Access, 2024.

[6] Zhao Y, Wang S, Duan H. LSPI: Heterogeneous Graph Neural Network Classification Aggregation Algorithm Based on Size Neighbor Path Identification[J]. arXiv preprint arXiv:2405.18933, 2024.

[7] Zhao Y, Xu S, Duan H. HGNN− BRFE: Heterogeneous Graph Neural Network Model Based on Region Feature Extraction[J]. Electronics, 2024, 13(22): 4447.

[8] Ergün, Esra, et al. "Modeling internet access at home by fractional calculus and a correlation analysis with human development index." TWMS Journal Of Applied And Engineering Mathematics (2024).